\documentclass{article}
\usepackage{spconf,amsmath,graphicx,amssymb, amsfonts, amsthm}
\usepackage[british]{babel}
\usepackage{setspace,epsfig,epstopdf,fleqn,latexsym}
\usepackage{fancyhdr, color}
\usepackage{tabulary}
\usepackage{multirow}
\usepackage{setspace}
\usepackage{indentfirst}
\usepackage{color,soul}
\usepackage{subfigure}
\title{Data-driven tight frame for Cryo-EM image denoising and conformational classification}
\name{Yin Xian*$^{\ddagger}$\thanks{Yin Xian and Hanlin Gu have equivalent contributions.}, Hanlin Gu*, Wei Wang$^{\dagger}$, Xuhui Huang$^{\dagger}$, Yuan Yao*, Yang Wang* and Jian-Feng Cai*}
\address{*Department of Mathematics, Hong Kong University of Science and Technology\\
$^{\dagger}$Department of Chemistry, Hong Kong University of Science and Technology \\
$^{\ddagger}$Department of Applied Mathematics, Beijing Computational Science Research Center}
\begin{document}
%\ninept
%
\maketitle
\begin{abstract}
The cryo-electron microscope (cryo-EM) is increasingly popular these years. It helps to uncover the biological structures and functions of macromolecules. In this paper, we address image denoising problem in cryo-EM. Denoising the cryo-EM images can help to distinguish different molecular conformations and improve three dimensional reconstruction resolution. We introduce the use of data-driven tight frame (DDTF) algorithm for cryo-EM image denoising. The DDTF algorithm is closely related to the dictionary learning. The advantage of DDTF algorithm is that it is computationally efficient, and can well identify the texture and shape of images without using large data samples. Experimental results on cryo-EM image denoising and conformational classification  demonstrate the power of DDTF algorithm for cryo-EM image denoising and classification.

\end{abstract}
\section{Introduction}
\label{sec:intro}
The cryo-electron microscope (cryo-EM) has been established as one of the fundamental techniques in structural biology. It can help to understand the macromolecules' structure, the arrangement of the atoms, and the biological mechanism of proteins~\cite{nogales2015development, cheng2015primer}. Unlike X-ray crystallography, cryo-EM does not require crystallization. Crystallization may change the conformation of the macromolecules, and many proteins and viruses are resistant to it~\cite{cheng2015single}. Cryo-EM is advantageous over Nuclear Magnetic Resonance spectroscopy (NMR) in solving macromolecules in native state.
%To prepare an cryo-EM image is also less computationally expensive than NMR~\cite{perilla2017cryoem}.
However, because of the limitation of resolution, cryo-EM was not popular in the past. Recent revolutionary advancement in detectors and softwares have improved the resolution of cryo-EM to atomic scale, and it is significantly popular these years~\cite{cheng2015single}. The Nobel Prize in Chemistry in 2017 was awarded for work that developed cryo-EM.

The cryo-EM images are created by the electron microscope that provides a top view of the molecules that are frozen in a thin layer of vitreous ice. The created image is called micrograph~\cite{wang2006cryo}. Image processing is crucial and it helps remove bad image samples, and facilitates orientation estimation, 3D inversion, 3D reconstruction and conformational classification~\cite{bhamre2017algorithms}.

%In order to have a Single Particle Reconstruction (SPR) using cryo-EM, that is, reconstruct a three dimensional molecular structure from the images, preprocessing the micrograph images is crucial. After processing the images, bad images will be removed, and the signal to noise ratio of the images is improved. It facilitates subsequent steps: orientation estimation and 3D inversion~\cite{bhamre2017algorithms}.

The challenge of cryo-EM images processing is that the images are highly influenced by noise~\cite{luvcic2013cryo, kuhlbrandt2014microscopy}. The point spread function of the microscope also blurs the images. The noise comes from various sources, and the type and level of noise in the dataset are unknown. The noise will obscure the conformational difference of molecules. It will also obscure projection of the same molecular structure  in different viewing directions. In this paper, we address the image denoising problem in cryo-EM, and evaluate the effect of noise reduction in 2D conformational classification.

A lot of methods have been proposed to remove noise in cryo-EM images. Singer and his group have designed a toolbox ASPIRE, and proposed Covariance Wiener Filtering (CWF)~\cite{bhamre2016denoising} for image denoising. CWF needs large samples of data in order to estimate the covariance matrix correctly, and have good denoising effect. They also proposed class averaging method, such as vector diffusion map~\cite{singer2012vector} for image denoising. These methods operate on Fourier domain. Other than that, the non-local mean method~\cite{darbon2008fast} has also been applied for cryo-EM image denoising.

In this paper, we propose to use the multi-image data-driven tight frame (DDTF)~\cite{cai2014data, bao2013fast} for cryo-EM image denoising. The DDTF method is inspired by the wavelet tight frame method~\cite{daubechies2003framelets} and the K-SVD method~\cite{elad2006image}.  It uses learned filters to form a tight frame. %The tight frame satisfies the perfect reconstruction property which is appealing to image processing task.
The Unitary Extension Principle (UEP) condition~\cite{shen2010wavelet} can be used to construct tight frames. However, it is not easy to satisfy the UEP condition. The DDTF algorithm relaxes the UEP condition, and generate filters with orthogonality. In the image patch space, the generated filters form an orthogonal dictionary. The K-SVD method needs a highly redundant dictionary to obtain a sparse code. The DDTF simplifies the process to obtain the filters coefficients, and reduce the computational cost compared with K-SVD. The use of data-driven tight frame can also better represents images with rich textures compared with standard wavelet methods and PDE based methods~\cite{cai2014data}.

We further use the denoised images for conformational classification. Experimental results show that DDTF outperform BM3D and KSVD in image denoising and classification. It demonstrates the power of DDTF method for cryo-EM image processing.

\section{Background}
\label{sec:background}
The problem of cryo-EM image formation model is:
\cite{frank2006three}:
\begin{align}
G(u)=C*Y(u)+Z(u)
\end{align}
where $Y$ be the clean ideal image, and $Z$ be the additive noise. Let $C$ be the point spread function of the microscope. $G$ is the measure image in real space, and $*$ is the convolution operator. The Fourier transform of the point spread function is the Contrast Transfer Function (CTF). In order to obtain an image close to the original image, the level of noise has to reduce, and the point spread function effect needs to be estimated. In this paper, we are focusing on reducing the noise level of the cryo-EM images. When the point spread function is known, the estimated image can be obtained by deconvolving the denoised images with the point spread function.

Singer and his group have developed Covariance Wiener Filtering (CWF) for cryo-EM image denoising. The procedure of this method is to first estimate the covariance matrix of the clean images from the noisy images, and then apply the traditional wiener filtering method, with the use of the estimated covariance matrix, for denoising~\cite{bhamre2016denoising}. The drawback of this method is that it needs a large number of images to accurately estimate the covariance matrix. When the covariance matrix is not accurately estimated, the performance of this method is not good. In this paper, we are seeking to denoise a single cryo-EM image well when the number of images is limited.

In the image processing area, the Block-matching and 3D filtering (BM3D) method~\cite{dabov2007image} is considered as an effective baseline. It groups similar and nonlocal image patches into a 3D array and filters the 3D array.  The image patches are then put back to the original positions and reweighed to form a denoised image. BM3D works particularly well for images with self-similarities.

K-SVD~\cite{elad2006image} is a dictionary learning method. K-SVD embeds the local overcompleted dictionary into a global Bayesian estimator. For a given noisy image $G$, the formula for K-SVD image denoising is to solve:
\begin{align}
\hat{\alpha}=\arg\min\limits_{\alpha}||D\alpha-G||_2^2+\mu||\alpha||_0
\label{eq:ksvd}
\end{align}
where $D$ is the dictionary, and $\alpha$ is the sparse code. The denoised image is given by $\hat{Y}=D\hat{\alpha}$.
The algorithm introduces the idea of updating image representation adaptively, and iteratively updates the sparse coding step and the dictionary update step.  Because the dictionary in K-SVD is unstructured, the computational cost of this method is heavy.

%Total variation~\cite{rudin1992nonlinear} (TV) is an effective filtering method for smoothing noise while preserving edges of the images. The Bayesian model is applied to the image denoising problem. It forms a minimization problem:
%\begin{align*}
%\min\limits_{Y\in L^2(\Omega)} \lambda F(Y)+\frac{1}{2}\int_{\Omega} |Y(u)-X(u)|^2du
%\end{align*}
%where $F(Y)$ is a total variation norm of the image pixels. $\lambda$ is a weight, and $Y$ is a $L^2(\Omega)$ function that is square integrable. The extension of TV model is Mumford-Shah model.

%recovering piecewise constant signals.

%The transform domain method uses orthogonal basis such as wavelets, curvelets and contourlets for image representation. BM3D finds the image

%The spatial domain methods include local and nonlocal filters, which use the similarities between pixels or patches in images for representation. The transform domain methods include BM3D

%The wavelet tight frame method generate an operator $W$ that consists of a set of predefined framelet filters $\{a_i\}_{i=1}^m$. $\{a_i\}_{i=1}^m$ forms a tight frame in $l_2(\mathbb{Z})$ and satisfies the unitary extension principle (UEP)~\cite{}. The filter set $\{a_i\}_{i=1}^m$, in image processing, defines a low pass, a band-pass and a high pass filters~\cite{}. Besides wavelet tight frame method, ridgelet, curvelet and shearlet tight frame have also been proposed, and they are particularly applicable to cartoon-type images.

\section{Multi-image Data-driven tight frame}
\label{sec:ddtf}
The data-driven tight frame (DDTF) is proposed based on the wavelet tight frame method and the K-SVD method. Compared with the wavelet tight frame, as well as the ridgelet, curvelet and shearlet tight frame methods, the data adpative tight frame method is effective to process natural images that are rich with texture~\cite{cai2014data}.

% which are particularly applicable to cartoon-type images,
%It generates an operator $W$ that consists of a set of framelet filters $\{a_i\}_{i=1}^m$. $\{a_i\}_{i=1}^m$ that form a tight frame in $l_2(\mathbb{Z})$ and satisfies the unitary extension principle (UEP)~\cite{}. The filter set $\{a_i\}_{i=1}^m$, in image processing, defines a low pass and a high pass filters~\cite{}.

\noindent \textbf{DDTF process:} Given an image $G$ of size $m\times n$, let $W$ be the analysis operator, its adjoint $W^T$ is a synthesis operator defined by filters $\{a_i\}_{i=1}^{r}$:
$W^T=[\mathcal{S}_{a_1}, \mathcal{S}_{a_2}, \cdots, \mathcal{S}_{a_r}]$,
where $\mathcal{S}_{a}$ is a linear convolution operator:~\cite{cai2014data}
\begin{align*}
[S_a v](n)=[a*v](n)=\sum\limits_{k\in \mathbb{Z}} a(n-k) v(k),
\end{align*}
where $v$ and $a$ are in $l_2$ space. $\mathcal{S}_a$ is of size $L\times L$, where $L=m\times n$. The column of $W^T$ forms a tight frame, and
\begin{align}
W^TW=I_L.
\label{eq:uep}
\end{align}
$I_L$ is an identity matrix of size $L$. A tight frame $W^T$ can be constructed by the minimization~\cite{cai2014data}:
\begin{align}
\small
&\min\limits_{\alpha, \{a_i\}_{i=1}^{r}}||\alpha-W(a_1,a_2,\cdots, a_{r})G ||_2^2 +\lambda^2||\alpha||_0
\end{align}
The filter coefficients $\{a_i\}_{i=1}^{r}$ and the sparse frame coefficients $\alpha$ can be solved iteratively. Specifically, given initial filter $\{a_i^{(0)}\}_{i=1}^r$,  at the step $k$, $k=1,2,\cdots, M$, we have,
\begin{align}
\small
&\alpha^{(k)}:=\min\limits_{\alpha}||\alpha-W(a_1^{(k)},a_2^{(k)},\cdots, a_{r}^{(k)})G||_2^2 +\lambda^2||\alpha||_0 ~\label{eq:alpha}\\
&\{a_i^{(k+1)}\}_{i=1}^{r}:=\min\limits_{\{a_i\}_{i=1}^{r}}||\alpha^{k}-W(a_1,a_2,\cdots, a_{r})G||_2^2~\label{eq:filter}
\end{align}
For eq.~(\ref{eq:alpha}), the filter coefficients $\{a_i\}_{i=1}^{r}$ are given, and $\alpha$ is update. For eq.~(\ref{eq:filter}), the sparse frame coefficient $\alpha$ is given, and  $\{a_i\}_{i=1}^{r}$ is update.

It can be proved that $\alpha^{*}=T_{\mu}(WG)$ is a unique solution for eq.~(\ref{eq:alpha}), where $T_{\mu}$ is a hard thresholding operator.  Let $A=[\hat{a}_1, \hat{a}_2, \cdots, \hat{a}_r]$, where $\hat{a}$ is a vectorized form of a 2D filter $a$. The unique solution of eq.~(\ref{eq:filter}) can be obtained by $A^{*}=\frac{1}{r}QP^T$. $P$ and $Q$ satisfies the singular value decomposition of $\bar{G}\bar{\alpha}$, that is $\bar{\alpha} \bar{G}=PDQ^T$.
$\bar{G}=[g_1, \cdots, g_n]$, where $g_i$ is the $i$-th vectorized image patches of $G$.
 and $\bar{\alpha}$ is the corresponding sparse frame coefficient matrix~\cite{cai2014data}. We can get the filter coefficients $\{a_i^{(0)}\}_{i=1}^r$ by $A^{*}$. The denoised image can be obtained by
\begin{align}
G^{*}=W^T(T_{\mu}(WG)).
\label{eq:frame_denoise}
\end{align}

\noindent \textbf{Connection with dictionary learning:} This subsection shows the DDTF in the image patch space is essentially dictionary learning.  Let $k_1\times k_2$ as the image patch size. For an image $G$, let $X=[x_{1}, x_{2}, \cdots, x_{mn}]$ is the image patch for $G$. $x_{j}$ is the $j$-th vectorized patch in $G$. Considering a sparse approximation for the dataset $X$. the frame operator $W$ becomes an orthogonal dictionary. The computation of the algorithm can be further accelerated. Since every signal has both high frequency and low frequency components, represent the orthogonal dictionary as $W=[A_1, A_2]$, where $A_1$ is a predefined low pass filter, and $A_2$ is the learned hight pass filters.  Eq.~(\ref{eq:alpha}) and eq.~(\ref{eq:filter}) then becomes~\cite{bao2013fast}
\begin{align}
&\min\limits_{A_2, \alpha} ||X-[A_1,A_2] \alpha ||_2^2 +\lambda^2||\alpha||_0 \\
&\min\limits_{A_2}||X-(A_1\alpha_{A_1}+A_2\alpha_{A_2})||_2^2
\end{align}
where $\alpha=[\alpha_{A_1},\alpha_{A_2}]$ denotes the codes associated with $A_1$ and $A_2$. The orthogonality constraint of W, according to eq.~(\ref{eq:uep}), becomes
\begin{align}
A_2^TA_2=I_r; A_1^TA_2=0.
\end{align}

The procedure to update the $A_2$ and $\alpha$ is similar to the tight frame case. The denoise dimage patches can be also obtained by the formula
\begin{align}
X^{*}=WT_{\mu}(W^TX),
\label{eq:denoise}
\end{align}
where $W=[A_1, A_2^{*}]$ in this case. Re-synthesizing the image patches, we can obtain the denoised image.

\noindent \textbf{Multi-image denoising:} For multi-image denoising, $W$ is learned from multiple images.  Given $N$ input images $\{G_i\}_{i=1}^{N}$ of size $m\times n$, take $k_1\times k_2$ as the image patch size as before, and concatenate the image patches together to form the input:
\begin{align*}
X=[X_1, X_2, \cdots, X_N] \in \mathbb{R}^{k_1k_2\times Nmn} .
\end{align*}
where $X_i=[x_{i,1}, x_{i,2}, \cdots, x_{i,mn}]$ is the image patch for the $i$-th image $G_i$. Apply eq.~(\ref{eq:alpha}) and eq.~(\ref{eq:filter}) to update and obtain the sparse frame coefficient $\alpha$ and filter coefficients $\{a_i\}_{i=1}^{r}$. Apply eq.~(\ref{eq:denoise}) to reduce the noise level of the image patches, and re-synthesize to obtain the denosied images.

The advantage of using multiple images is that using multiple images can better capture the distribution of noise information and perform denoising. The position of the macromolecules is not always at the center of the cryo-EM images. Constructing $W$ from mutiple images can better eliminate the background noise.

%by applying the alternating iteration scheme to find $A_2$ and $\alpha_{A_2}$

 %Let $\alpha=[\alpha_{A_1},\alpha_{A_2}]$ denotes the codes associated with $A_1$ and $A_2$. $P_{A_1}$ be the projection operator that $P_{A_1}v=A_1(A_1^Tv)$, where $v\in \mathbb{R}^n$. The dictionary update $A_2$ can be obtained through the following model:
%\begin{align*}
%&\min\limits_{A_2\in\mathbb{R}^{n\times r}}||X-(A_1\alpha_{A_1}+A_2\alpha_{A_2})||_F^2 \\
%&\text{s.t.}~~A_2^TA_2=I_r; A_1^TA_2=0
%\end{align*}
%It has a unique solution given by $A_2^{*}=PQ^T$, where $P$ and $Q$ satisfies the singular value decomposition:
%\begin{align*}
%(I_n-P_{A_1})X\alpha_{A_2}^T=PDQ^T
%\end{align*}

%The process of image denoising is by applying the alternating iteration scheme to find $A_2$ and $\alpha_{A_2}$

\section{Experimental Results}
\label{sec:experiment}

\subsection{Datasets}
\label{ssec:data}
We test the efficiency of the denoising algorithms on the structurally heterogeneous synthetic dataset. The dataset is generated based on five representative atomistic structures of Thermus aquaticus RNA polymerase. It is obtained from the snapshots of molecular dynamics simulations. The images are generated by two dimensional projections of a three dimensional model of RNA polymerase. The size of the images are $128\times 128$.

We evaluate the effectiveness of the noise reduction algorithms according to the Peak Signal to Noise Ratio (PSNR), and conformational classification error rate.  The point spread function is set to be a Delta function in this case. Additive Gaussian noise is added to the clean images. The clean images are obtained from the voxelization and 2D projection of the atomic structure of the molecules. It is prepared by the Xmipp package~\cite{marabini1996xmipp}. The noisy datasets are prepared at different level of signal to noise ratio (SNR). The SNR is defined by
$\frac{E_s}{\sigma^2}$, where $E_s$ is the power of the signal, and $\sigma^2$ is the variance of the noise.

\subsection{Denoised results}
\label{ssec:denoised}
\begin{figure}[!ht]
\centering
%\hfill
\subfigure[Clean image]{\includegraphics[width=.325\linewidth]{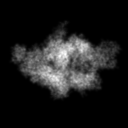}}
%\hfill
\subfigure[Noisy image]{\includegraphics[width=.325\linewidth]{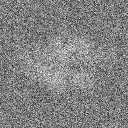}}
%\hfill
\subfigure[CWF]{\includegraphics[width=.325\linewidth]{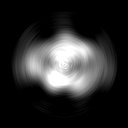}}
\\
\centering
\subfigure[DDTF]{\includegraphics[width=.325\linewidth]{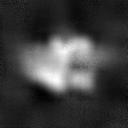}}
%\subfigure[TV]{\includegraphics[width=.325\linewidth]{01_TV}}
\subfigure[BM3D]{\includegraphics[width=.325\linewidth]{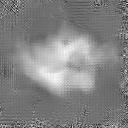}}
%\hfill
\subfigure[K-SVD]{\includegraphics[width=.325\linewidth]{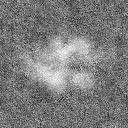}}
%\hfill
\caption{Noise reduction of different algorithms when SNR=0.1}
\label{fig:img_denoise}
\end{figure}

We first of all evaluate the CWF algorithm on the noisy images when SNR is 0.1. The result is shown in Figure~\ref{fig:img_denoise} (c). Compared with the results shown in~\cite{bhamre2016denoising}, which uses 10,000 images to estimate the mean and covariance matrix for denoising, we use 2031 images for experiment given the limitation of computational resource. According to the results, CWF fails to capture the shape and content of the polymerase, and is not suitable for the conformational classification in this case. It is possible that the estimated covariance matrix is not accurately estimated and leads to not desirable denoising result.

We evaluate DDTF, BM3D and K-SVD on the dataset. The levels of SNR are 0.8, 0.4, 0.2, 0.1, 0.05 and 0.01. There are 2031 images for each SNR level. The pixel value of each image is in the range of 0 to 255. In the experiments, the image patch size of DDTF, BM3D and K-SVD are 16. We use 20 images to obtain the filter coefficients and sparse code of the multi-image DDTF. The initial filters are generated by a discrete cosine function.  For K-SVD, we use random initial dictionary for initialization.

Figure~\ref{fig:img_denoise} shows the  noise reduction effect of of each algorithms when SNR is equal to 0.1. Table~\ref{table:1_psnr} shows the average PSNR. For an image $\textbf{x}$ of size $L\times M$, the PSNR of its estimated image $\hat{\textbf{x}}$ is defined by
\begin{align*}
PSNR(\textbf{x},\hat{\textbf{x}})=10\log_{10}\frac{255^2}{\frac{1}{LM}\sum\limits_{i=1}^L\sum\limits_{j=1}^M(\hat{\textbf{x}}(i,j)-\textbf{x}(i,j))^2} .
\end{align*}

%\begin{table}[h!]
%%\small
%\caption{Average MSE} %\label{table1}%title of the table
%\label{table:1_img}
%\vspace{1pt}
%\centering % centering table
%\begin{tabular}{cccc}
%\hline\hline
%SNR  & DDTF  & BM3D & K-SVD\\
%\hline
%\\ [0.01ex]
%0.8 & \textbf{0.0077}  & 0.0225 & 0.0421    \\
%0.4 &  \textbf{0.0208}  & 0.0536 & 0.069 \\
%0.2 & \textbf{0.0198}  & 0.1093 & 0.1051\\
%0.1 & \textbf{0.0367}  & 0.1685 & 0.1439 \\
%0.05 & \textbf{0.0479}  & 0.1958 & 0.1738 \\
%0.01 & \textbf{0.1125}  & 0.1982 & 0.1999
%\\ [1ex]
%\hline %inserts single line
%\end{tabular}
% is used to refer this table in the text
%\end{table}

\begin{table}[h!]
%\small
\caption{PSNR (dB)} %\label{table1}%title of the table
\label{table:1_psnr}
%\vspace{1pt}
\centering % centering table
\begin{tabular}{cccc}
\hline\hline
SNR  & DDTF  & BM3D & K-SVD\\
\hline
%\\ [0.01ex]
0.8 & \textbf{45.2005}{\scriptsize$\pm 1.51$}  & 40.5436{\scriptsize$\pm 1.57$} & 37.8226{\scriptsize$\pm 0.72$}    \\
0.4 &  \textbf{42.4517}{\scriptsize$\pm 1.61$}  & 36.7738{\scriptsize$\pm 1.50$} & 35.6769{\scriptsize$\pm 0.66$} \\
0.2 & \textbf{41.3008}{\scriptsize$\pm 1.66$} & 33.6792{\scriptsize$\pm 1.34$} & 33.8494{\scriptsize$\pm 0.60$}\\
0.1 & \textbf{38.9199}{\scriptsize$\pm 1.69$}  & 31.7994{\scriptsize$\pm 1.07$} & 32.4848{\scriptsize$\pm 0.53$} \\
0.05 & \textbf{36.6697}{\scriptsize$\pm 1.46$}  & 31.1473{\scriptsize$\pm 0.85$} & 31.6649{\scriptsize$\pm 0.52$} \\
0.01 & \textbf{33.2232}{\scriptsize$\pm 1.35$}  & 31.0944{\scriptsize$\pm 0.77$} & 31.0573{\scriptsize$\pm 0.60$}
\\ [1ex]
\hline %inserts single line
\end{tabular}
% is used to refer this table in the text
\end{table}

The PSNR measures the ratio of the maximum possible power of a signal and the power of noise that affects the fidelity of signal representation. It is generally used to show the image reconstruction quality. According to the results shown in  Table~\ref{table:1_psnr}, DDTF performs better than BM3D and K-SVD in PSNR. From Figure~\ref{fig:img_denoise}, the multi-image DDTF better preserves the shape, and has less artifacts compared with other methods. Since the cryo-EM images have little self-similarity pattern, the BM3D method, which uses nonlocal information of the images for denoising, does not perform well.

\subsection{Conformational classification results}
\label{ssec:classification}
\begin{figure}[!ht]
% \hfill
\centering
\subfigure[Clamp-open conformation]{\includegraphics[width=.43\linewidth,height=2.5cm]{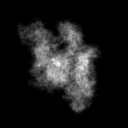}}
 \hfill
\subfigure[Clamp-close conformation]{\includegraphics[width=.43\linewidth,height=2.5cm]{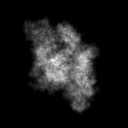}}
\caption{Illustration of conformations of RNA polymerase}
\label{fig:class}
\end{figure}

After image denoising, we classify the "clamp-open" structure and "clamp-close" structure of the RNA polymerase among the images. Classification helps us to solve the relative population distribution of stable conformations of macromolecules. The illustration of these two structures are shown in Figure~\ref{fig:class}. The numbers of "clamp-open" structure and "clamp-close" structure images are 420 and 429 respectively.

We compared the denoised image with DDTF, BM3D and K-SVD methods, and noisy image without denoising as inputs for the conformational classification. We perform a brute force classification of images by pixel information. We select 31 template images from landmarks around the "North Pole" of the sphere for each conformation. Because the molecules can be rotated at random angels, we rotate and reflect the template images, and the Euclidean distance is calculated between the tested images and the rotated and reflected templates. Classification is performed based on these distances. When the distance of the noisy image is close to the "clamp-open" template, the image is classified as "clamp-open" class. Similarly, when the noisy image is close to "clamp-close" template, the image is in the "clamp-close" class.

\begin{table}[h!]
%\small
\caption{Classification Error Rate (\%)} %\label{table1}%title of the table
\label{table:class}
\vspace{2pt}
\centering % centering table
\begin{tabular}{ccccc}
\hline\hline
SNR  & DDTF  & BM3D & K-SVD & Noisy Image\\
\hline
%\\ [0.01ex]
0.8 & \textbf{0}  & 0 & 0 & 0.59   \\
0.4 &  \textbf{0} & 0.35 & 0 & 22.85\\
0.2 & \textbf{0.35}  & 0.12 & 0.58 & 48.00\\
0.1 & \textbf{0.47}  & 10.95 & 14.72  & 50\\
0.05 & \textbf{2} & 36.04 & 44.76 & 50
%0.01 & \textbf{25.8} & 49
\\ [1ex]
\hline %inserts single line
\end{tabular}
% is used to refer this table in the text
\end{table}

The classification results are shown in Table~\ref{table:class}. According to the results, we can see that DDTF performs better than BM3D and K-SVD in classification error rate. The lower the SNR, the better DDTF compared with BM3D and KSVD. DDTF can well capture the shape and texture of images. It helps to distinguish different molecules conformation.

\section{Conclusion}
\label{sec:conclud}
In this paper, we applied DDTF method for cryo-EM image denoising and conformational classification. The denoising effect of data-driven tight frame is better than K-SVD and BM3D. It also improves conformational classification accuracy over other algorithms. Our research demonstrates that data-driven tight frame is an effective algorithm for cryo-EM image processing.

\vfill\pagebreak

%\section{Acknowledgement}
%\label{sec:refs}
%We thank Professor Xuhui Huang from Department of Chemistry at Hong Kong University of Science and Technology for providing the datasets.

\bibliographystyle{IEEEbib}
\bibliography{strings,refs}

\end{document}